\documentclass[aps,prl,showpacs]{revtex4}
\usepackage{amsmath}
\usepackage{graphicx}
\usepackage{bm}

\begin{document}

\title{Dirac sextic oscillator in the constant magnetic field}
\date{\today}
\author{Ramazan Ko\c{c}}
\email{koc@gantep.edu.tr}
\affiliation{Department of Physics, Faculty of Engineering University of Gaziantep, 27310
Gaziantep, Turkey}

\author{Mehmet Koca}
\email{kocam@squ.edu.om}
\affiliation{Department of Physics, College of Science, Sultan Qaboos University, PO Box
36, Al-Khod 123, Muscat, Sultanate of Oman}

\begin{abstract}
We introduce a Dirac equation which reproduces the usual radial
sextic oscillator potential in the non-relativistic limit. We
determine its energy spectrum in the presence of the magnetic
field. It is shown that the equation is solved in the context of
quasi-exactly-solvable problems. The equation possesses hidden
$sl_{2}$-algebra and the destroyed symmetry of the equation can be
recovered for a specific values of the magnetic field which leads
to exact determination of the eigenvalues.
\end{abstract}



\section{Introduction}

A dirac equation with an interaction linear in coordinates was considered
long ago\cite{ito} and recently rediscovered in the context of the
relativistic many body theories\cite{mosh1}. The equation is named \textit{%
Dirac oscillator}, since in the non-relativistic limit it becomes a harmonic
oscillator with a very strong spin-orbit coupling term. Dirac oscillator has
attracted much attention and the concept gave rise to a large number of
papers concerned with its different aspects\cite%
{moreno,mosh2,ferk,ros,ho,mart,levai,alhaidari}.

During the last years two-dimensional electron systems have become an active
research subject due to the rapid growth in nanofabrication technology that
has made possible to the production of low dimensional structures like
quantum wells, quantum wires, quantum dots etc.\cite{villa1,villa2,chak}. In
non-relativistic case, the two dimensional parabolic potential has often
been used to describe the spectrum of the electron in confined
two-dimensional systems. For the relativistic case, the spectrum and
properties of the such systems can be determined by using two dimensional
Dirac oscillator\cite{villa2,schak,naagu}. Despite their simplicity, both
relativistic and non-relativistic oscillator potentials appear to be a good
approximation to complicated low dimensional nanostructures.

In order to present a more realistic model we construct deformed form of the
Dirac oscillator including a term $qr^{3}$. The Dirac oscillator with a
deformed term becomes radial sextic oscillator potential in the
non-relativistic limit. The non-relativistic sextic oscillator potential is
quasi-exactly-solvable(QES) for which it is possible to determine
algebrically a part of spectrum but not whole spectrum\cite%
{turb,bender1,bender2}. It will be shown that the solution of the
relativistic sextic oscilator can also be treated in the context of the QES
problem.

We also investigate the effect of the magnetic field on the Dirac sextic
oscillator. It will be shown that for a specific values of the magnetic
field the Dirac sextic oscillator problem is exactly solvable.

The paper is organized as follows. In section 2 we discuss the construction
of the Dirac sextic oscillator in polar coordinate, in the presence of the
magnetic field. In section 3, we solve the Dirac sextic oscillator in the
context QES problem. We show that the problem possesses hidden $sl_{2}-$%
algebra. In section 4 we briefly conclude our method and results.

\section{Construction of the Dirac sextic oscillator}

The $(2+1)-$dimensional Dirac equation for free particle of mass $M$ in
terms of two-component spinors $\psi ,$ can be written as%
\begin{equation}
E\psi =\left[ \sum_{i=1}^{2}c\beta \gamma _{i}p_{i}\mathbf{+}\beta Mc^{2}%
\right] \psi  \label{e1}
\end{equation}%
Since we are using only two component spinors, the matrices $\beta $ and $%
\beta \gamma _{i}$ are conveniently defined in terms of the Pauli spin
matrices which satisfy the relation $\sigma _{i}\sigma _{j}=\delta
_{ij}+\varepsilon _{ijk}\sigma _{k},$ given by
\begin{equation}
\beta \gamma _{1}=\sigma _{1};\quad \beta \gamma _{2}=\sigma _{2};\quad
\beta =\sigma _{3}.  \label{e2}
\end{equation}%
In $(2+1)-$dimensions, the momentum operator $p_{i}$ is two component
differential operator, $\mathbf{p}=-i\hbar (\partial _{x},\partial _{y})$
for free particle$.$ In the presence of the magnetic field it is replaced by
$\mathbf{p}\rightarrow \mathbf{p-}e\mathbf{A}$, where $\mathbf{A}$ is the
vector potential, and the 2D Dirac oscillator can be constructed by changing
the momentum $\mathbf{p}\rightarrow \mathbf{p-}im\omega \beta \mathbf{r}$.
We are now looking for some expressions on the right hand side of (\ref{e1})
that can be interrepted as radial sextic oscillator Hamiltonian in the
non-relativistic limit. For this purpose we introduce the following momentum
operator%
\begin{equation}
\mathbf{p}\rightarrow \mathbf{p-}i\sigma _{3}(m\omega -qr^{2})\mathbf{r}
\label{e3}
\end{equation}%
where $q$ is a constant and the Dirac equation takes the form%
\begin{eqnarray}
\left[ E-\sigma _{0}mc^{2}\right] \psi &=&c\left[ \sigma
_{+}(p_{x}-ip_{y}-i(M\omega -qr^{2})(x-iy))\right] \psi +  \nonumber \\
&&c\left[ \sigma _{-}(p_{x}+ip_{y}-i(M\omega -qr^{2})(x+iy))\right] \psi
\label{e4}
\end{eqnarray}%
In polar coordinate, $x=rcos\phi ,\quad y=rcos\phi $, the 2D Dirac equation (%
\ref{e4}) can be written as%
\begin{eqnarray}
&&\varepsilon ^{2}\psi _{1}=-c^{2}\hbar ^{2}\left[ \frac{\partial ^{2}}{%
\partial r^{2}}+\frac{1}{r}\frac{\partial }{\partial r}+\frac{1}{r^{2}}\frac{%
\partial ^{2}}{\partial \phi ^{2}}+\frac{2i\left( M\omega -qr^{2}\right) }{%
\hbar }\frac{\partial }{\partial \phi }\right] \psi _{1}  \nonumber \\
&&+c^{2}\left( q^{2}r^{6}-2M\omega qr^{4}+\left( M^{2}\omega ^{2}-4\hbar
q\right) r^{2}+2\hbar M\omega \right) \psi _{1}  \label{e5}
\end{eqnarray}%
where $\psi _{1}=\psi _{1}(r,\phi )$ is the upper component of the spinor $%
\psi $ and $\varepsilon ^{2}=E^{2}-M^{2}c^{4}$. The substitution of%
\begin{equation}
\psi _{1}(r,\phi )=\frac{e^{-im\phi }}{\sqrt{r}}f(r)  \label{e6}
\end{equation}%
leads to the following equation%
\begin{equation}
\varepsilon ^{2}f(r)=-c^{2}\hbar ^{2}\frac{\partial ^{2}f(r)}{\partial r^{2}}%
+c^{2}\left[ V(r)+2\hbar M\omega (1-m)\right] f(r)  \label{e7}
\end{equation}%
where $V(r)$ is given by%
\begin{equation}
V(r)=\frac{\hbar ^{2}\left( m^{2}-\frac{1}{4}\right) }{r^{2}}%
+q^{2}r^{6}-2M\omega qr^{4}+\left( M^{2}\omega ^{2}-2\hbar q(2-m)\right)
r^{2}  \label{e7a}
\end{equation}%
It is not difficult to see that in the non-relativistic limit (\ref{e7})
corresponds to the Schr\"{o}dinger equation with radial sextic oscillator
potential with a spin-orbit coupling term. In the presence of the symmetric
gauge vector potential $A(x,y)=B/2(-y,x,0)$, the potential (\ref{e7a}) takes
the form%
\begin{eqnarray}
V(r) &=&\frac{\hbar ^{2}\left( m^{2}-\frac{1}{4}\right) }{r^{2}}+\hbar
eB(m-1)+  \nonumber \\
&&q^{2}r^{6}-\left( 2M\omega -eB\right) qr^{4}+\left( \left( M\omega -\frac{%
eB}{2}\right) ^{2}-2\hbar q(2-m)\right) r^{2}  \label{e7b}
\end{eqnarray}%
From now on we restrict ourselves to the solution of (\ref{e7}) and (\ref%
{e7a}).

\section{Method of Solution}

In this section we show that the Dirac sextic oscillator (\ref{e7}) is one
of the recently discovered quasi-exactly solvable operator \cite%
{turb,bender1}. It is well known that the underlying idea behind the quasi
exact solvability is the existence of a hidden algebraic structure. Let us
introduce the following realization of the $sl_{2}-$algebra:
\begin{equation}
J_{+}=\rho ^{2}\frac{d}{d\rho }-j\rho ,\quad J_{-}=\frac{d}{d\rho },\quad
J_{0}=\rho \frac{d}{d\rho }-\frac{j}{2}  \label{e9}
\end{equation}%
The generators satisfy the commutation relations of the $sl_{2}-$algebra for
any value of the parameter $j$. If $j$ is a positive integer the algebra (%
\ref{e9}) possesses $j+1-$dimensional irreducible representation:
\begin{equation}
P_{j+1}=\left\langle 1,\rho ,\rho ^{2},\ldots ,\rho ^{j}\right\rangle
\label{e10}
\end{equation}%
The linear and bilinear combinations of the operators given in (\ref{e9})
are quasi exactly solvable, when the space is defined in (\ref{e10}). In
order to show the Dirac sextic oscillator has a $sl_{2}$-symmetry, let us
consider the following combinations of the operators (\ref{e9}):
\begin{equation}
T=-J_{0}J_{-}+\frac{j+2}{2}J_{-}+16c^{4}\hbar ^{3}qJ_{+}+4c^{2}\hbar M\omega
J_{0}.  \label{e11}
\end{equation}%
Then, the eigenvalue problem can be written as
\begin{equation}
TP_{k}(\rho )=\left( \varepsilon ^{2}+2Mc^{2}\hbar \omega \right) P_{k}(\rho
)  \label{e12}
\end{equation}%
where $P_{k}(\rho )$ is the $k^{th}$ degree polynomial in $\rho $. Let us
turn our attention to the Dirac sextic oscillator (\ref{e7}). Introducing a
new function
\begin{equation}
f(r)=r^{m-\frac{1}{2}}e^{-\frac{m\omega }{2\hbar }r^{2}-\frac{q}{4\hbar }%
}F(r)  \label{e13}
\end{equation}%
and then changing the variable $r=2c\hbar \sqrt{\rho },$ we obtain the
following expression%
\begin{eqnarray}
&&\varepsilon ^{2}F(\rho )=  \nonumber \\
&&-\rho \frac{\partial ^{2}F(\rho )}{\partial \rho ^{2}}+\left(
m-1+4c^{2}\hbar \omega M\rho -16qc^{4}\hbar ^{3}\rho ^{2}\right) \frac{%
\partial F(\rho )}{\partial \rho }  \nonumber \\
&&-c^{2}\left( 4M\hbar \omega (m-1)-16qc^{2}\hbar ^{3}(m-2)\rho \right)
F(\rho ).  \label{e13a}
\end{eqnarray}%
We can show that the eigenvalue equation (\ref{e7}) and (\ref{e12}) are
identical, when the following holds:%
\begin{equation}
m=j+2,\quad F(\rho )=P_{k}(\rho ).  \label{e14}
\end{equation}%
When the generators act on the polynomial (\ref{e10}), we can obtain the
following recurrence relation
\begin{eqnarray}
&&16qc^{4}\hbar ^{3}(k-j)P_{k+1}(\varepsilon )+(\varepsilon
^{2}+4Mc^{2}\hbar \omega (j-k+1))P_{k}(\varepsilon )-  \nonumber \\
&&k(j-k+2)P_{k-1}(\lambda )=0  \label{e15}
\end{eqnarray}%
with the initial condition $P_{0}(1)=1$. If $\varepsilon _{i}$ is a root of
the polynomial $P_{k+1}(\varepsilon )$, the wavefunction is truncated at $%
k=j $ and belongs to the spectrum of the Hamiltonian $T$. This property
implies that the wavefunction is itself the generating function of the
energy polynomials. The roots of the recurrence relation (\ref{e15}) can be
computed and the first few of them are given by%
\begin{eqnarray}
P_{1}(\varepsilon ) &=&\varepsilon ^{2}+4Mc^{2}\hbar \omega  \nonumber \\
P_{2}(\varepsilon ) &=&\left( \varepsilon ^{2}+4Mc^{2}\hbar \omega \right)
\left( \varepsilon ^{2}+8Mc^{2}\hbar \omega \right) -32qc^{4}\hbar ^{3}
\nonumber \\
P_{3}(\varepsilon ) &=&\varepsilon ^{6}+8c^{2}\hbar (3\varepsilon
^{4}M\omega +48c^{4}\hbar ^{2}M\omega (M^{2}\omega ^{2}-3q\hbar )+
\label{e16} \\
&&2c^{2}\hbar \varepsilon ^{2}(11M^{2}\omega ^{2}-10q\hbar ))  \nonumber \\
P_{4}(\varepsilon ) &=&\varepsilon ^{8}+40\varepsilon ^{6}Mc^{2}\hbar \omega
+80\varepsilon ^{4}c^{4}\hbar ^{2}(7M^{2}\omega ^{2}-6\hbar q)+  \nonumber \\
&&128\varepsilon ^{2}c^{6}\hbar ^{3}M\omega (25M^{2}\omega ^{2}-69\hbar
q)+6144c^{8}\hbar ^{4}(3\hbar ^{2}q^{2}-6M^{2}\omega ^{2}\hbar q+M^{4}\omega
^{4}).  \nonumber
\end{eqnarray}%
The function $P_{j}(\rho )$ forms a basis for $sl_{2}-$algebra and it can be
written in the form
\begin{equation}
P_{j}(\rho )=\sum\limits_{k=0}^{j}c_{k}P_{k}(\varepsilon )\rho ^{k}.
\label{e17}
\end{equation}%
In the presence of the magnetic field the Hamiltonian can be solved by the
same procedure given above. When the magnetic field $B=2M\omega /e,$ the
Hamiltonian (\ref{e7}) takes the form%
\begin{eqnarray}
&&\varepsilon ^{2}f(r)=-c^{2}\hbar ^{2}\frac{\partial ^{2}f(r)}{\partial
r^{2}}+  \nonumber \\
&&c^{2}\left[ \frac{\hbar ^{2}\left( m^{2}-\frac{1}{4}\right) }{r^{2}}%
+q^{2}r^{6}-2\hbar q(2-m)r^{2}+4\hbar M\omega (1-m)\right] f(r)  \label{e17a}
\end{eqnarray}%
We define the wavefunction%
\begin{equation}
f(r)=r^{\frac{1}{2}-m}e^{-\frac{qr^{4}}{4\hbar }}F(r)  \label{e18}
\end{equation}%
and changing the variable $r=2c\hbar \sqrt{\rho },$ we obtain the following
differential equation%
\[
\varepsilon ^{2}F(\rho )=-\rho \frac{d^{2}}{d\rho ^{2}}+\left(
j+1-16qc^{4}\hbar ^{3}\rho ^{2}\right) \frac{\partial F(\rho )}{\partial
\rho }+16jqc^{4}\hbar ^{3}\rho F(\rho ).
\]%
When $F(\rho )=P_{k}(\rho ),$where $P_{k}(\rho )$ is $k^{th}$ degree
polynomial in $\rho ,$ we obtain the following reccurence relation%
\[
16qc^{4}\hbar ^{3}P_{k+1}(\varepsilon )-\varepsilon ^{2}P_{k}(\varepsilon
)+k(j+2-k)P_{k-1}(\varepsilon )=0
\]%
The polynomial $P_{k}(\varepsilon )$ vanishes for $k=j+1$ and the roots of $%
P_{j}(\varepsilon )$ belongs to the spectrum of the (\ref{e17a}). We list
the first ten of them below%
\begin{eqnarray}
P_{1}(\varepsilon ) &=&\varepsilon ^{2}  \nonumber \\
P_{2}(\varepsilon ) &=&\varepsilon ^{4}-2\eta ^{2}  \nonumber \\
P_{3}(\varepsilon ) &=&\varepsilon ^{6}-10\varepsilon ^{2}\eta ^{2}
\nonumber \\
P_{4}(\varepsilon ) &=&\varepsilon ^{8}-30\varepsilon ^{4}\eta ^{2}+72\eta
^{4}  \nonumber \\
P_{5}(\varepsilon ) &=&\varepsilon ^{10}-70\varepsilon ^{6}\eta
^{2}+712\varepsilon ^{2}\eta ^{4}  \label{e19} \\
P_{6}(\varepsilon ) &=&\varepsilon ^{12}-140\varepsilon ^{8}\eta
^{2}+3820\varepsilon ^{4}\eta ^{4}-10800\eta ^{6}  \nonumber \\
P_{7}(\varepsilon ) &=&\varepsilon ^{14}-252\varepsilon ^{10}\eta
^{2}+14796\varepsilon ^{6}\eta ^{4}-164592\varepsilon ^{2}\eta ^{6}
\nonumber \\
P_{8}(\varepsilon ) &=&\varepsilon ^{16}-420\varepsilon ^{12}\eta
^{2}+46380\varepsilon ^{8}\eta ^{4}-1307600\varepsilon ^{4}\eta
^{6}+4233600\eta ^{8}  \nonumber \\
P_{9}(\varepsilon ) &=&\varepsilon ^{18}-660\varepsilon ^{14}\eta
^{2}+125004\varepsilon ^{10}\eta ^{4}-7250320\varepsilon ^{6}\eta
^{6}+88504707\varepsilon ^{2}\eta ^{8}  \nonumber
\end{eqnarray}%
where $\eta ^{2}=16qc^{4}\hbar ^{3}.$ Therefore we can obtain the
eigenfunction of (\ref{e17a}) in the closed form. We conclude that
anhormonic interaction destroys the general symmetry of the Dirac equation,
but the specified magnetic field can restore the symmetries of the Dirac
equation. This feature implies that for the specific values of the magnetic
field $B=2\hbar \omega /e,$ analytical solutions of the roots of the
polynomials are avaliable.

\section{Conclusion}

Due to the interest of the lower dimensional field theory and condensed
matter physics, we have constructed Dirac sextic oscillator in
two-dimensional space in the presence of the magnetic field. We have given
eigenstates of the corresponding equation interms of the orthogonal
polynomials\cite{kraz}. It has been shown that the Dirac sextic oscillator
possesses hidden $sl_{2}-$symmetry.

We note that similar constructions of the other quasi-exactly solvable Dirac
equations also seem possible by considering the momentum operator of the
form: $\mathbf{p}\rightarrow \mathbf{p-}i\sigma _{3}(m\omega +v(r))\mathbf{r.%
}$ The method we have introduced can be applied to construct Dirac equation
including the other non-relativistic potentials such as P\"{o}schle-Teller,
Eckart, Scarf, Hult\.{e}n, etc. We hope that the Dirac sextic oscillator may
be used as a model in the related fields of the physics.


\begin{thebibliography}{99}
\bibitem{ito} D. It\^{o}, K. Mori and E. Carriere, \textit{Nuovo Cimento A}
\textbf{51} (1967) 1119.

\bibitem{mosh1} M. Moshinsky and A. Szczepaniak, \textit{J. Phys. A: Math.
Gen.} \textbf{22 }(1989) L817.

\bibitem{moreno} M. Moreno and A. Zentella, \textit{J. Phys. A: Math. Gen}.
\textbf{22} (1989) L821.

\bibitem{mosh2} M. Moshinsky, and Y. F. Smitnov, \textit{The Harmonic
Oscillator in Modern Physics}, (Harwood Academic Publishers, Amsterdam,
1996).

\bibitem{ferk} N. Ferous and A. Bounames, \textit{Phys.Lett.A} \textbf{325}
(2004) 21.

\bibitem{ros} P. Rojmez, R. and Arvieu, \textit{J. Phys. A: Math. Gen.}
\textbf{32} (1999) 5367.

\bibitem{ho} C. L. Ho and P. Roy, \textit{Ann. Phys(N.Y.)} \textbf{312}
(2004) 161.

\bibitem{mart} R. Mart\'{\i}nez, M. Moreno and A. Zentella, \textit{Rev.
Mex. F\'{\i}s}. \textbf{36} (1990) S176.

\bibitem{levai} G. Levai and A. Del Sol Mesa, \textit{J. Phys. A: Math. Gen.}
\textbf{29} (1996) 2827.

\bibitem{alhaidari} A. D. Alhaidari, \textit{J. Phys. A: Mat. Gen.} \textbf{%
34} (2001) 9827.

\bibitem{villa1} V. M. Villalba and R. Pino, \textit{Mod. Phys. Lett. B}
\textbf{17} (2003) 1331.

\bibitem{villa2} V. M. Villalba, \textit{Phys. Rev. A} \textbf{49} (1994)
586.

\bibitem{chak} A. M. J. Schakel, \textit{Phys. Rev. D} \textbf{43} (1991)
1428.

\bibitem{schak} A. M. J. Schakel and G. W. Semenoff, \textit{Phys. Rev. Lett.%
} \textbf{66} (1991) 2653.

\bibitem{naagu} A. Neagu and A. M. J. Schakel, \textit{Phys. Rev. D} \textbf{%
48} (1993) 1785.

\bibitem{turb} A. V. Turbiner and A. G. Ushveridze, \textit{Phys Lett. A}
\textbf{126}, (1987) 81.

\bibitem{bender1} C. M. Bender and M. Moshe, \textit{Phys. Rev. A} \textbf{55%
} (1997) 2625.

\bibitem{bender2} C. M. Bender and S. Boettcher, \textit{J. Phys. A: Math.
Gen.} \textbf{31} (1998) L273.

\bibitem{kraz} A. Krajewska, A. Ushveridze and Z. Walczak, \textit{Mod.
Phys. Lett. A} \textbf{12} (1997) 1131.
\end{thebibliography}
\end{document}